\documentclass[conference,a4paper,final]{IEEEtran}
\usepackage{amsmath}
\usepackage{amssymb}
\usepackage{graphicx}
\usepackage{epsf}
\usepackage{epstopdf}
\usepackage{paralist}
\usepackage {eqnarray}
\usepackage{cite}

\newcommand\blfootnote[1]{%
  \begingroup
  \renewcommand\thefootnote{}\footnote{#1}%
  \addtocounter{footnote}{-1}%
  \endgroup
}
\begin{document}
%
\title{Effect of Local Binding on Stochastic Transport in Ion Channels}

\author{\IEEEauthorblockN{
I. Kh. Kaufman\IEEEauthorrefmark{1},
W. A. T. Gibby\IEEEauthorrefmark{1},
D. G. Luchinsky\IEEEauthorrefmark{1}\IEEEauthorrefmark{2},
P. V. E. McClintock\IEEEauthorrefmark{1},
}
\IEEEauthorblockA{\IEEEauthorrefmark{1}Department of Physics, Lancaster University, Lancaster LA1 4YB, UK\\Email: p.v.e.mcclintock@lancaster.ac.uk}
\IEEEauthorblockA{\IEEEauthorrefmark{2}SGT Inc., Greenbelt MD, 20770, USA}
}
\maketitle
\begin{abstract}
Ionic Coulomb blockade is an electrostatic phenomenon recently discovered in low-capacitance ion channels/nanopores. Depending on the fixed charge that is present, Coulomb blockade strongly and selectively influences the ease with which a given type of ion can permeate the pore. The phenomenon arises from the discreteness of the charge-carriers 
and it manifests itself strongly for divalent ions (e.g.\ Ca$^{2+}$). Ionic Coulomb blockade is closely analogous to electronic Coulomb blockade in quantum dots. 
In addition to the non-local 1D Coulomb interaction considered in the standard Coulomb blockade approach, we now propose a correction to take account of the singular part of the attraction  to the binding site (i.e.\ local site binding). We show that this correction leads to a geometry-dependent shift of the single-ion barrierless resonant conduction points M$_0$. We also show that local ion-ion repulsion accounts for a splitting of Ca$^{2+}$ profiles observed earlier in Brownian dynamics simulations.
\end{abstract}
\section{Introduction}
\label{sec:intro}
Ion channels provide for the  selective transport of physiologically important ions (e.g.\ Na$^+$, K$^+$ and Ca$^{2+}$) through the bilipid membranes of biological cells. The channels consist of nanopores through proteins embedded in the membrane. Their selectivity for particular ions is determined by the electrostatically-driven stochastic motion of ions within a short, narrow  selectivity filter (SF) carrying a binding site with fixed negative charge $Q_f$.

The permeation of ions through the pore is governed by ionic Coulomb blockade (ICB), a phenomenon that manifests itself in low-capacitance systems. It arises as a consequence of the discreteness of the charge-carriers, the dielectric self-energy $U_q^{SE}$, an electrostatic exclusion principle, and sequential pore neutralisation as additional ions enter the pore \cite{Krems:13,Kaufman:15,Kaufman:15b}. ICB manifests itself strongly for divalent ions (e.g.\ Ca$^{2+}$) \cite{Kaufman:15}. ICB is closely similar to its electronic counterpart in quantum dots \cite{Beenakker:91}.
\begin{figure}[t]
\begin{center}
\includegraphics[width=1.0\linewidth]{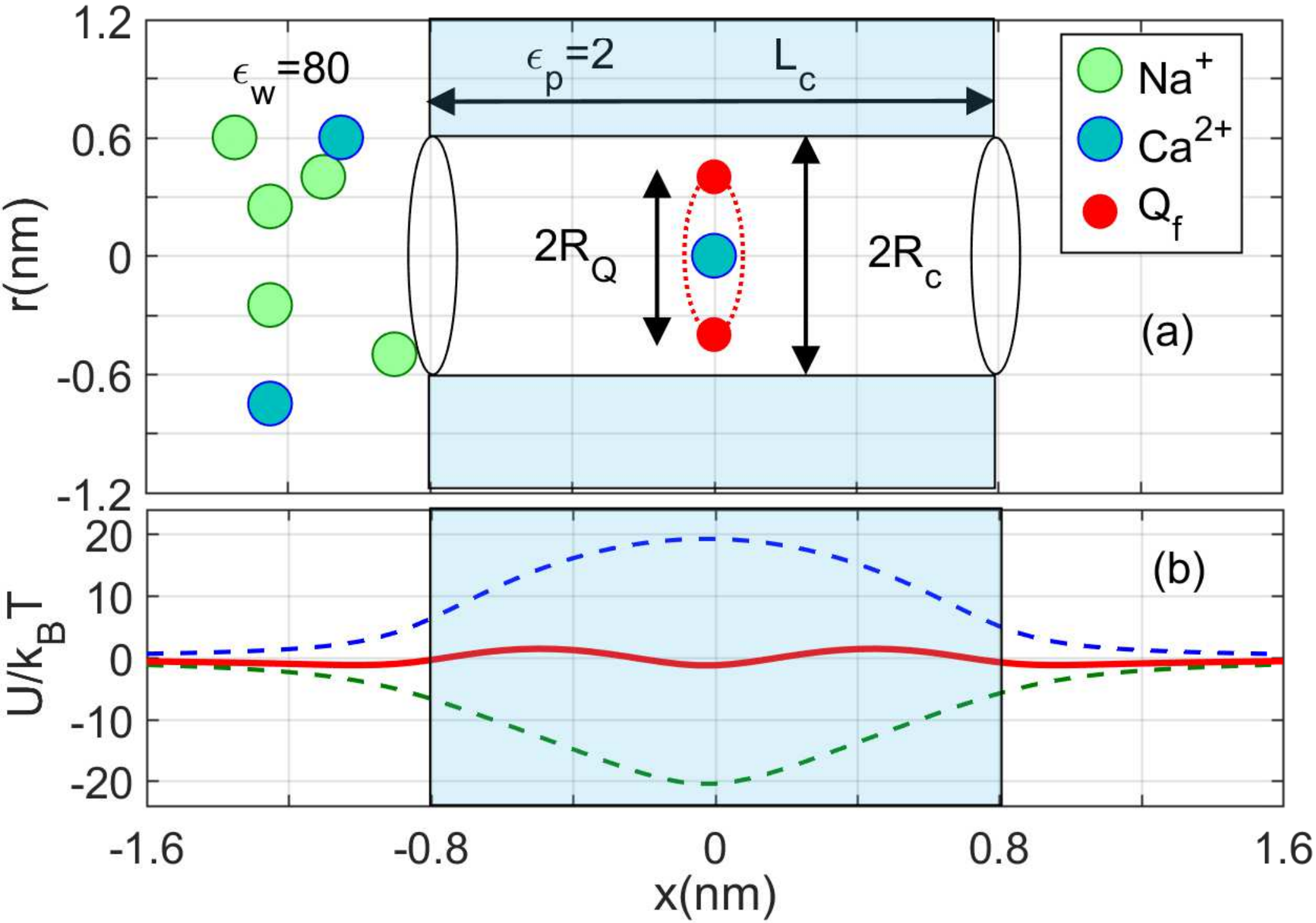}
\end{center}
\caption{ (Modified from \cite{Kaufman:15b}) Extended electrostatic model of the selectivity filter of Ca$^{2+}$ or Na$^+$ channel. (a) The model represents a channel as a negatively-charged, axisymmetric,  water-filled, cylindrical pore of radius $R_c=0.3-0.5$nm and length $L_c=1.2-1.6$nm through the protein hub in the cellular membrane.  The $x$-axis is coincident with the channel axis and $x=0$ in the center of channel. There is a uniformly-charged, rigid ring of negative charge $|Q_f|=(0-8)e$ of radius $R_Q$. Ions move in single file along the channel axis.  (b) Energetics of moving Ca$^{2+}$ ion for fixed charge $Q_f=-1e$ ( by Brownian dynamics simulations). The dielectric self-energy barrier $U_q^{SE}$ (dashed blue line) is balanced by the site attraction $U_{qQ}^{CB}$ (dashed green line) resulting in an almost barrier-less energy profile $U_b$ (red solid line).} \label{fig:channel}
\end{figure}

The basic ICB description of the permeation and selectivity of ion channels has already been presented \cite{Kaufman:15b}. 
Here we extend this basic model by the introduction of corrections to allow for the singular part of the attraction of ions to the binding site (i.e.\ local site binding), in %
addition to the non-local 1D Coulomb interaction considered in the ICB model \cite{Kaufman:15}. The geometry-dependent shift of the ICB calcium resonant point M$_0$ resulting from this correction leads to a changed threshold $IC_{50}$ for divalent blockade. We will also show that the presence of local (singular) ion-ion repulsion is what leads to the splitting of the Ca$^{2+}$ axial occupancy profile seen earlier in Brownian dynamics simulations \cite{Corry:01, Kaufman:13a}.

We start in Sec. \ref {sec:model} by description of extended electrostatic model of ion channels. 
In Sec. \ref {sec:icb} we briefly describe the ICB model of permeation and selectivity of calcium/sodium channel. Sec. \ref {sec:lb} introduces extension  accounting for a local binding, followed in Sec.  \ref {sec:repulsion} by consideration of local ion-ion repulsion and resulted multi-ion splitting of calcium profiles. Finally, in Sec. \ref{sec:concl} we summarize and draw conclusions.

In what follows, with SI units $\varepsilon_0$ is the permittivity of free space, $e$ is proton charge,  $k_B$ is Boltzmann's constant and $T$ is the temperature; (I)CB is (ionic) Coulomb blockade, BD is Brownian dynamics, LB is local binding, LR is local repulsion, SF is the selectivity filter and SE is the self-energy.
\blfootnote {The research was supported by the Engineering and Physical Sciences Research Council (EPSRC)
UK (grant No. EP/M015831)}
\pagebreak
\section {Extended Electrostatic Model of Ion Channel}
\label{sec:model}
%
\begin{figure}[t]
\begin{center}
\includegraphics[width=1.0\linewidth]{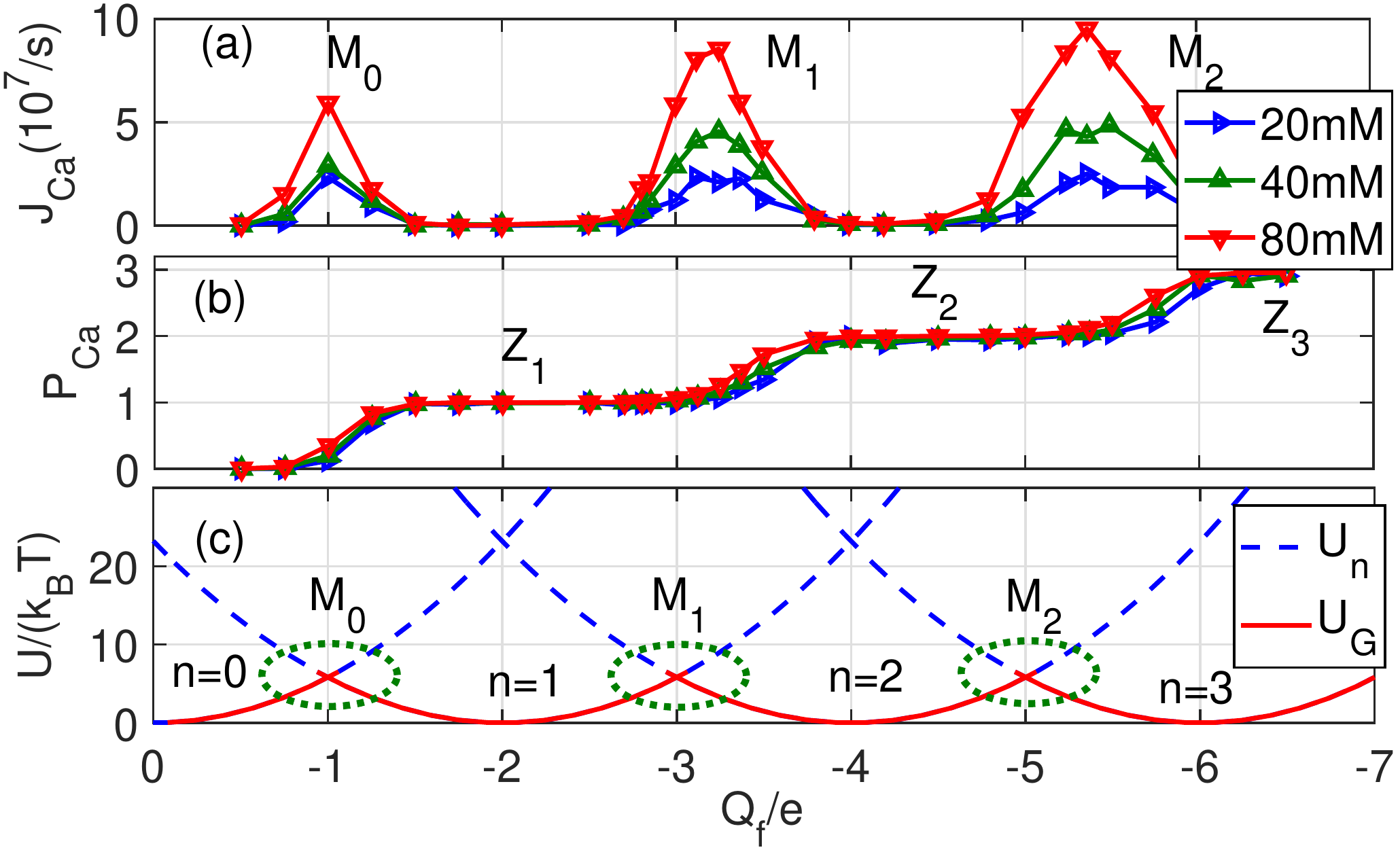}
\end{center}
\caption{(Reworked from \cite{Kaufman:15b}) Brownian dynamics simulations of multi-ion conduction and occupancy in a Ca$^{2+}$ channel model {\it vs} the effective fixed charge $Q_f$.
(a) Conduction bands in the Ca$^{2+}$ current $J$ for pure Ca$^{2+}$ baths of different concentration (20, 40 and 80mM as indicated). (b) Coulomb staircase for the occupancy $P$.
(c) The excess self-energy $U_n$ and ground state energy $U_G$ {\it vs} $Q_f$ for channels containing $n=0,1,2$ and 3 Ca$^{2+}$ ions.
The conduction bands M$_n$ 
and the blockade/neutralisation points Z$_n$ 
are discussed in the text.}
\label{fig:ca_bands}
\end{figure}
Fig.\ \ref{fig:channel}(a) shows the extended electrostatic model of the SF of a calcium/sodium ion channel, We represent it as a negatively-charged, axisymmetric, water-filled, cylindrical pore of radius $R_c=0.3-0.5$nm and length $L_c=1.2-1.6$nm ~through the protein hub in the cellular membrane  The $x$-axis is coincident with the channel axis and $x=0$ in the center of channel. 

There is a symmetrically-placed, uniformly-charged, rigid ring of negative charge $|Q_f|=(0-8)e$. Extending the earlier model, the radius of this charged ring $R_Q$ could be different from the channel radius $R_c$, corresponding to the charged residues moving partially into the channel. We take both the water and the protein to be homogeneous continua with dielectric constants $\varepsilon_w=80$ and $\varepsilon_p=2$, respectively. 

Fig.\ \ref{fig:channel}(b) illustrates the phenomenon of resonant barrier-less conduction, which is  typical of electrostatic models and which occurs  when the energy of ion-site attraction $U_{gQ}$ balances the dielectric self-energy barrier $U_{q}^{SE}$ \cite{Eisenman:83,Yesylevskyy:05,Kaufman:15}.

This generic electrostatic channel model is similar to that used previously \cite{Nonner:00,Corry:01,Zhang:05,Giri:11}. Details of the model, and its validity and limitations, have already been discussed \cite{Kaufman:13a}.
\section {Ionic Coulomb blockade}
\label{sec:icb}
We  consider the stochastic transport of a fully-hydrated Ca$^{2+}$ ion having charge $q=2e$. 

Fig.\ (\ref{fig:ca_bands})(a) illustrates the multi-ion Ca$^{2+}$ conduction bands and (b) shows the corresponding Coulomb staircase in occupancy, as revealed by Brownian dynamics simulations. Steps of this staircase are described by Fermi-Dirac function \cite {Kaufman:15,Kaufman:15b}.  The ground state energy diagram for such conductance is plotted in (c). The ICB model  \cite{Kaufman:15b} states that resonant (barrier-less) conduction points M$_n$  occur when the difference in free energy $\Delta G_{n}$ between states $s_{n+1}$ (with $n+1$ ions near the center of the SF) and $s_{n}$ (channel with $n$ ions in the SF plus 1 ion in the bulk) is zero:
\begin{equation}
\Delta G_{n}= G_{n+1} - G_{n} = U_{n+1}-U_{n}- T\Delta S_n
\end{equation}
where $G_i$ refers to state $s_i$, $U_n$ is the potential energy and $\Delta S_n$ is the entropy difference. When $\partial U_n/\partial Q_f =0$, there are stable ICB points Z$_n$.

The standard ICB model \cite {Kaufman:15} assumes that $U_n$ with \{n\} similar ions near the centre is equal to the dielectric self-energy $U_n^{SE}$ of the excess charge of the SF $Q_n=n q + Q_f$ :
\setlength{\arraycolsep}{0.1em}
\begin {equation}
U_{n} = U_{n}^{SE}=\frac{Q_n^2}{2C_s }; \quad
C_s=\frac{4 \pi \varepsilon_0 \varepsilon_w R_c^2}{ L_c};  
\label{equ:excess0}
\end {equation}
where $C_s$ is the SF self-capacity. 

For simplicity, we consider the first resonant point M$_0$ \cite{Kaufman:13a} corresponding to the movement of a single ion through an otherwise empty SF (the $s_0 \rightarrow s_1 \rightarrow s_0$ transition) so we temporarily ignore the ion-ion interaction term $U_{qq}$ (see Sec.\ \ref {sec:repulsion}).
Expanding the quadratic form in (\ref {equ:excess0}) gives us the following decomposition for ion-related part of potential energy $U_q$ (for $n=1$)
\begin {eqnarray}
U_q^{CB}&=&U_q^{SE} +U_{qQ}^{CB} ; \\ 
\label{equ:decomp}
U_q^{SE}&=& \frac{q^2}{2C_s}; \quad 
U_{qQ}^{CB}=\frac{q Q_f}{C_s};  
\end {eqnarray}
where $U_q^{SE}$ 
is the ion self-energy 
, and $U_{qQ}^{CB}$ is the 1D Coulomb ion-site attraction energy 
\cite{Zhang:05, Kamenev:06}.

The base position (without the entropy term) for resonant conduction is  defined 
by the condition for barrier-less motion $\Delta U_n=U_{n+1} -U_{n} =0$ \cite{Kaufman:15,Kaufman:15b}:
\begin {equation}
M_0^{CB}=-\frac{q}{2}; \quad Z_1^{CB}=-q;
\end {equation}
whereas Z$_n^{CB}$ corresponds to the $Q_n=0$ SF neutralisation condition.
Inclusion of the entropy term $T\Delta S$ 
leads to a concentration-related shift of the resonance point \cite{Kaufman:15}:
\begin {equation}
  \delta M_0^{TS}= C_s \frac{k_B T}{q}\log (P_b
  ); \quad P_b=\frac{n_b}{n_0};
\end {equation}
where P$_b$ is the equivalent bulk occupancy related to the SF volume $V_c=\pi R_c^2 L_c$, $n_b$ is the number density of selected ions in the bulk, and $n_{0}=1/V_c$ is the reference density; note that $\delta M_0^{TS}=0$ for $P_b=1$ (i.e. for $n_b=n_0$).  For a typical SF geometry condition ($P_b=1$)  corresponds to the concentration [Ca]$_0\approx $200mM/l. The dependence of the resonance point's position on $V_c$ coincides with both simulations and with earlier analytic results \cite{Boda:07b,Malasics:09}.
\section {Local binding}
\label{sec:lb}
Next we introduce a local binding (LB) correction by adding the  ion-site ($q \Leftrightarrow Q_f$)  3D $\varepsilon$-screened Coulomb interaction (see \cite{Luchinsky:07}) with energy $U_{qQ}^{LB}$ to the total ion potential energy $U_q$.
It leads to a geometry-dependent shift $\delta M_0^{LB}$ in the resonance point M$_0^{LB}$: 
\begin {eqnarray}
U_q^{CB+LB}&=&U_q^{CB}+U_{qQ}^{LB};\ \
U_{qQ}^{LB} = \frac{1}{4 \pi \varepsilon_0} \frac{q Q_f }{\varepsilon_w R_Q}; \\
\delta M_0^{LB}&=&- M_0^{CB}\frac{ \beta_c}{1+\beta_c}; \quad \beta_c=\frac{R_c^2}{R_Q L_c};
 \label {equ:delta1}
\end {eqnarray}
where $\beta_c$ is the dimensionless``SF shape ratio''. 
For an embedded charge ring ($R_Q=R_c$), $\beta_c$ reduces to $\beta_c=R_c/L_c$.
For typical geometries ($R_c=R_Q=$0.3nm , $L_c=1.5$nm) the correction for LB can be about 0.2.

In summary,
\begin{equation}
      M_0= M_0^{CB} +\delta M_0^{LB}+\delta M_0^{TS}
      \label {equ:opt}
\end {equation}
The ``shift-equation'' (\ref {equ:opt}) allows us to describe the whole range of ICB phenomenona, as embodied in different shifts of M$_0$ and their possible interference such as a divalent blockade and  its dependence on $Q_f$ \cite{Elinor:95,Kaufman:15}, or concentration-related shifts of the Coulomb staircase --
\begin{figure}[t]
\begin{center}
\includegraphics[width=1.0\linewidth]{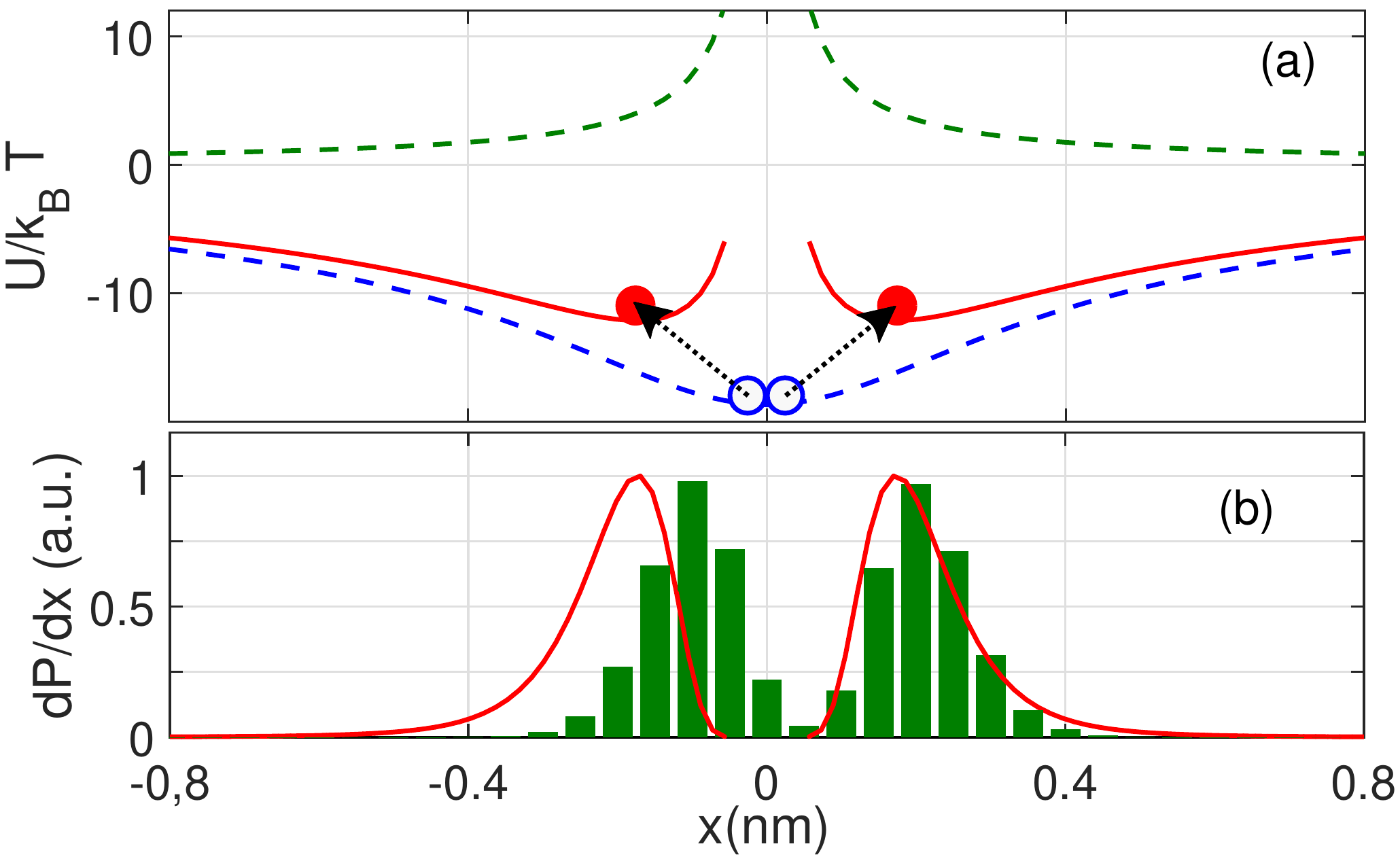}
\end{center}
\caption{Effect of local binding and local repulsion on the energy and occupancy profiles for Ca$^{2+}$ ions inside the selectivity filter, in accordance with (\ref {equ:profile}) for the Z$_1$ resonance point ($Q_f=4$) and $R_Q=R_c$. (a) 
Local binding component $U_{qQ}^{LB}$ (blue dashed line, open blue circles), local repulsion $U_{qq}^{LR}$ (green dashed line) and resulted energy profile $U_q$ (red solid line, red closed circles) for the split equilibrium positions. (b) The split calcium profile (green bars) obtained by self-consistent Brownian dynamics simulation \cite{Kaufman:13b} is consistent with the analytic result $\rho \propto \exp(-U_q/(k_B T))$ (red solid line)}
\label {fig:splitting}
\end{figure}
\begin {compactitem}
\item
The standard ICB optimal conduction point M$_0^{CB}$ defines the barrier-less point for ion of charge $q$ with $\beta_c \rightarrow 0$ at the standard bulk density $n_0$ ($P_b=1$).
\item
The LB shift $\delta M_0^{LB}$ accounts for the real shape of the SF and/or for $R_Q$.
\item
The concentration-related shift $\delta M_0^{TS}$ describes the influence of $n_b$.
\item
The interference between $\delta M_0^{LB}$ and $\delta M_0^{TS}$ could explain observable differences in divalent blockade thresholds $IC_{50}$ between equally-charged (D/E) mutants of the calcium \cite{Elinor:95} or bacterial sodium  channels \cite{Naylor:16}.
\end{compactitem}
The axial potential energy profiles  for the ICB model can be derived  from the 1D Coulomb gas approximation \cite{Zhang:05,Kamenev:06}:
\begin{eqnarray}
U_q^{SE}(x)&=&U_q^{SE}\left(1-(2x/L_c)^2\right) \\
U_{qQ}^{CB}(x)&=&U_{qQ}^{CB} \left(1-(2|x|/L_c)\right)  
\label {equ:profile}
\end{eqnarray}
where 
$U_q^{SE}(x)$ is the dielectric self-energy profile and $U_{qQ}^{CB}(x)$ is the ion-site binding energy.
Profile for LB correction can be calculated from Coulomb's law \cite{Luchinsky:07}:
\begin {eqnarray}
U_{qQ}^{LB}(x)&=&  U_{qQ}^{LB} \left(1+(x/R_Q)^2\right)^{-1/2}
\label{equ:U_LB}
\end{eqnarray}
\section {Local ion-ion repulsion}
\label {sec:repulsion}
In the above analytics we ignored ion-ion repulsion, but here we take it explicitly into account and work out its consequences. For simplicity we consider 2 similar ions located symmetrically ($-x_q, +x_q$) around $Q_f$ at $x=0$. In such a case, the 1D Coulomb ion-ion repulsion $U_{qq}^{CB}$ and the additional local 3D Coulomb repulsion $U_{qq}^{LR}$ are respectively:
\begin{equation}
U_{qq}^{CB} = -\frac{1}{4 \pi\varepsilon_0}
\frac{q^2 4x_q}{\varepsilon_w R_c^2}; 
\quad
U_{qq}^{LR}= \frac{1}{4 \pi\varepsilon_0}
\frac{q^2}{\varepsilon_w 2x_q}
\label {equ: int_ij}
\end{equation}
The total energy of an ion inside the SF, allowing for both LB and LR, can be expressed as:
\begin {equation}
U_{q}=U_{q}^{CB+LB}+\frac{1}{2}
(U_{qq}^{CB}+U_{qq}^{LR})
\end{equation}
Hence for the neutralization point Z$_n$ $Q_n=0 \Rightarrow U_n=0$, the energy profile is defined by local components only.

Fig. \ref{fig:splitting}(a) shows that the energy function $U_q(x)$ calculated according to (\ref {equ: int_ij}) has two symmetrical off-center minima $\pm x_{min}$ defining a splitting of the Ca$^{2+}$ occupancy profile. 
 Thus, by taking account of LB and of the local ion-ion repulsion in the SF, we arrive at an explicit, self-consistent, analytic explanation of the splitting of the multi-ion occupancy profiles observed earlier in Brownian dynamics simulations \cite{Corry:01, Kaufman:13a, Kaufman:13b}, in self-consistent numerical  solutions of the Poisson equation \cite {Kaufman:13b}, and in analytic non-self-consistent calculations \cite{Kitzing:92,Kharkyanen:10}. 
 This splitting leads to significant increases in ionic energy (as indicated by arrows) and eventually to knock-on escape. Note that the ``binding points'' ($\pm x_{min}$)  are unconnected with any physical binding sites different from the main $Q_f$-related site. This situation can be described as 
 ``'virtual sites' or self-organisation of ions inside the SF (see also \cite{Giri:11,Luchinsky:09a}) .

Fig. \ref{fig:splitting}(b) compares the Ca$^{2+}$ occupancy profile $\rho(x)$ returned by BD simulations (green histogram) with that estimated from the potential energy profile $U_q(x)$ as $\rho(x) \propto \exp(-U_q(x)/(k_B T))$ (red solid line). The satisfactory agreement obtained can be regarded as confirming the consistency of our model.

\begin{figure}[t]
\begin{center}
\includegraphics[width=1.0\linewidth]{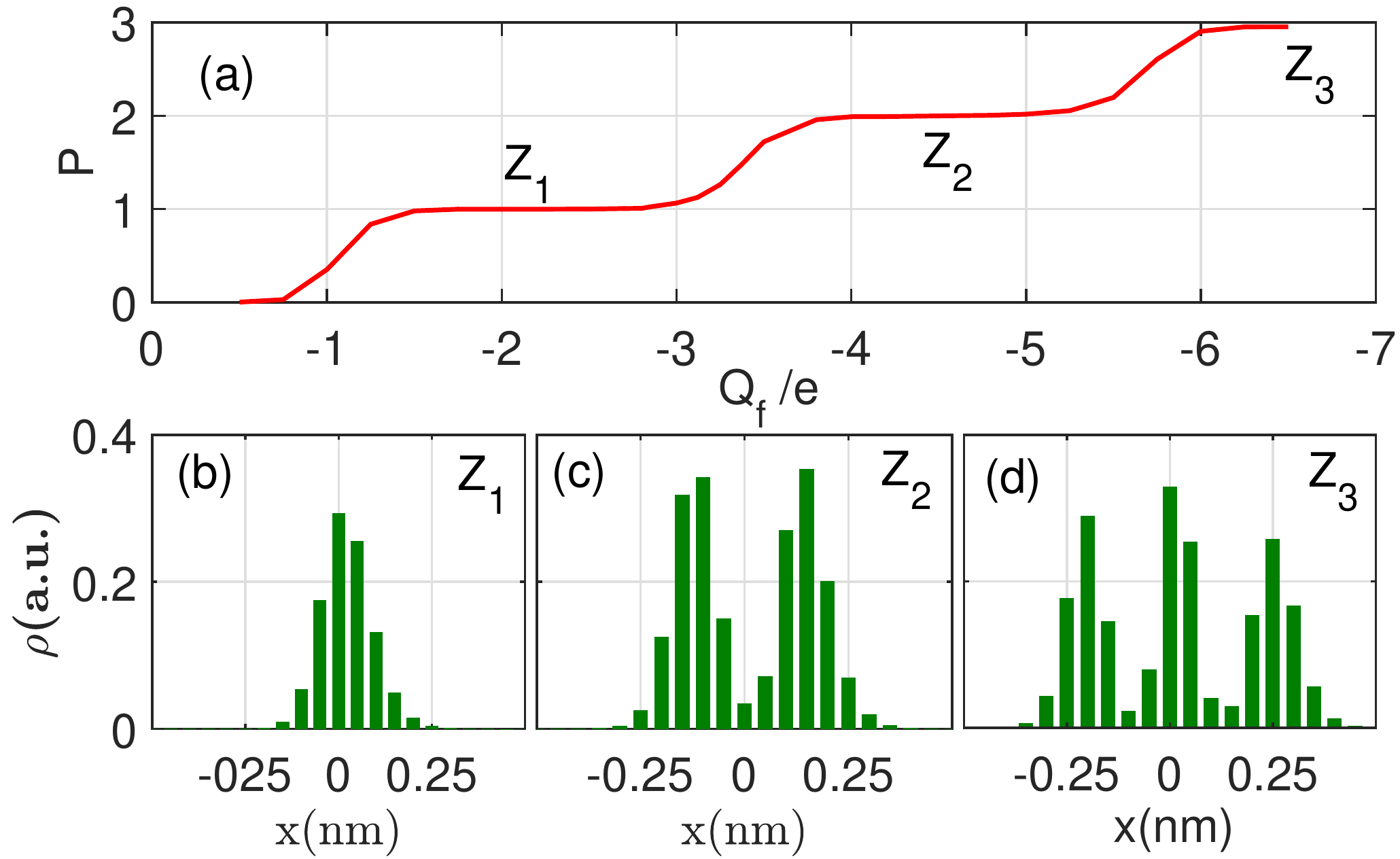}
\end{center}
\caption{Evolution of calcium profile {\it vs.} fixed charge $Q_f$ from Brownian dynamics simulations \cite {Kaufman:13b}. (a) Simulated calcium occupancy $P$ of the selectivity filter {\it vs} $Q_f$ (red solid line) for bath concentration [Ca]=80mM. Single-ion Z$_1$, double-ion Z$_2$ and triple-ion Z$_3$ neutralisation points are labeled. (b) Single-ion (unsplit) occupancy profile (green bars) for Z$_1$ point. (c) Double-ion (split) occupancy profile (green bars) for  Z$_2$ point. (d) Triple-ion (split) occupancy profile (green bars) for  Z$_3$ point. }
\label {fig:combo}
\end{figure}

Fig. \ref{fig:combo} shows the evolution of the Ca$^{2+}$ profile with varying $Q_f$ as found from BD simulations \cite {Kaufman:13b, Kaufman:13c}.  

Fig. \ref{fig:combo}(a) shows the simulated calcium occupancy $P$ of the SF vs $Q_f$ for bath concentration [Ca]=80mM (see Fig.\ref {fig:channel}(b)), demonstrating the standard Coulomb staircase shape with Fermi-Dirac steps. The single-ion Z$_1\approx2e$, double-ion Z$_2\approx 4.5e$ and triple-ion Z$_2\approx 6.5e$ blockade points  are indicated.

Fig. \ref{fig:combo} (b)-(d) represent BD-simulated profiles for different $Q_f$ values. In (b) there is an (unsplit) single-ion occupancy profile for the Z$_1$ point, which appears due to ICB and LB of the ion to $Q_f$. 

The local repulsion provides for self-organisation of the ions and splitting of the occupancy profile for the double-ion Z$_2$ point (c) and triple-ion Z$_3$ point (d).  Note that, for intermediate values of $Q_f$ (e.g. between Z$_1$ and Z$_2$), the \{n\}-state will be mixed (time shared), providing averaged profiles having an arbitrary number of peaks.
%
\section{Conclusions}
\label{sec:concl}
The effect of local binding on conduction and occupancy (analytic result (\ref {equ:opt})) is found to agree with Brownian dynamics simulations, thus putatively accounting for how the position of the resonant point M$_0$ is influenced by the radius of the Glutamate/Aspartate ring in NaChBac channels and their mutants \cite{kaufman:16b, Naylor:16, Guardiani:16}.  

The local binding and local repulsion lead to corrections of about 20$k_B T$ and to an observable splitting of the Ca$^{2+}$ occupancy profiles. The splitting of the Ca$^{2+}$ profile is found in reasonable quantitative agreement with the results of BD simulations \cite{Kaufman:13a}. 

These results are also applicable to artificial sub-nm pores \cite{Feng:16}.
\section* {Acknowledgment}
We are grateful to M.\ Di Ventra, R.\ S.\ Eisenberg, O.\ A.\ Fedorenko, C. \ Guardiani, S.\ K.\ Roberts, and A.\ Stefanovska %
for comments and useful discussions.
\bibliographystyle{IEEEtran}
\bibliography{ionchannels-}

\end{document}